\documentclass{emulateapj}

\usepackage{amsmath}
\usepackage{cases}
\usepackage{txfonts}
\usepackage{tipa}
\usepackage{amsfonts}
\usepackage{amssymb}
\usepackage{graphicx}
\usepackage{multirow}
\usepackage{float}
\usepackage{graphicx}
\usepackage{epstopdf}
\usepackage{multirow}
\usepackage{verbatim}
\usepackage{color}

\begin{document}

\title{Metallicity and Kinematics of the Galactic halo from the LAMOST sample stars}

\author{Shuai Liu\altaffilmark{1,2}, Cuihua Du\altaffilmark{1}, Heidi Jo Newberg\altaffilmark{3},
Yuqin Chen\altaffilmark{2}, Zhenyu Wu\altaffilmark{2,1}, Jun Ma\altaffilmark{2,1}, Xu Zhou\altaffilmark{2}, Zihuang Cao\altaffilmark{2}, Yonghui Hou\altaffilmark{4}, Yuefei Wang \altaffilmark{4} and Yong Zhang \altaffilmark{4}}

\affil{$^{1}$College of Astronomy and Space Sciences, University of Chinese Academy of Sciences, Beijing 100049, China; ducuihua@ucas.ac.cn \\
$^{2}$Key Laboratory of Optical Astronomy, National Astronomical Observatories, Chinese Academy of Sciences, Beijing 100012, China\\
$^{3}$Department of Physics, Applied Physics and Astronomy, Rensselaer Polytechnic Institute, Troy, NY 12180, USA\\
$^{4}$Nanjing Institute of Astronomical Optics $\&$ Technology, National Astronomical Observatories, Chinese Academy of Sciences, Nanjing 210042, China\\
 }

\begin{abstract}

\par We study the metallicity distribution and kinematic properties of 4,680 A/F/G/K-type giant stars with $|z|>$ 5 kpc selected from the LAMOST spectroscopic survey.  The metallicity distribution of giant stars with 5 $<|z|\leqslant$ 15 kpc can be described by a three-peak Gaussian model with peaks at [Fe/H] $\sim-0.6\pm0.1$, $-1.2\pm0.3$ and $-2.0\pm0.2$, corresponding to the ratio of 19$\%$, 74$\%$ and 7$\%$, respectively. The $\rm{[\alpha/Fe]}$ is used to associate the three peaks with the thick disk, inner-halo and outer-halo components of the Galaxy. 
The metallicity distribution of these giant stars, which is fit with Gaussians corresponding to the three components, show a growing fraction of inner-halo component and declining fraction of the thick-disk component with increasing distance from the Galactic plane.  
Adopting a galaxy potential model,  we also derive the orbital parameters of the sample stars,  such as orbit eccentricity and rotation velocity.
The peak values of derived orbital eccentricity for stars covering different metallicity regions maintain $e\sim$ 0.75, independent of height above the plane, within the range 5$<|z|<$ 15 kpc.  By comparing the MDFs of stars in different rotation velocity intervals, we find that the majority of the retrograde stars are more metal-poor than the prograde stars.

\end{abstract}

\keywords{Galaxy:abundance-Galaxy:halo-Galaxy:kinematics and dynamics}

\section{Introduction}

\par  The Galactic halo comprises only $1\%$ of the total stellar mass in our galaxy.  However, due to the very old age ($\sim$ 10 -13 Gyr) of most halo stars, they play a pivotal role in our study of the formation and evolution of our galaxy. 
In our current understanding of the evolutionary process that formed the halo stars, the different components of the halo have different formation mechanisms: (1) the accreted halo stars found in the outer halo formed mainly through the accretion of smaller stellar systems like dwarf spheroidal (dSph) galaxies \citep{Sales07,Diemand08,Springel08,Klypin11}; (2) the ``in situ'' halo stars primarily formed from dissipative collapse of gaseous material onto the central region of the Galaxy \citep{Zolotov10, Font11,Cooper15}, and were then joined by stars formed in the disk component and then kicked into the halo.  The proposed mechanisms for ejecting disk stars into the halo include dwarf galaxy heating \citep{Purcell10}, binary supernova ejection \citep{Bromley09} and other gravitational mechanisms.
 
\par Massive accretion events are believed to heat more metal-rich disk stars so that they are ejected into the halo \citep{Purcell10}. This theory is supported by recent observations.  For example, \cite{Hawkins15a} found a metal-rich halo star in the RAVE spectroscopic survey that has likely been dynamically ejected into the halo from the Galactic thick disk. Using the first Gaia data release, metal-rich halo stars have been discovered in the local stellar halo \citep{Bonaca17}. However, up until now the vast majority of stars found in the halo are the metal-poor stars, with metal abundances less than 1/10th of the solar value \citep{Chiba00, Carollo07,Carollo10}. 

In recent decades, evidence for the dual halo (the inner-halo and outer-halo populations) has been found by \cite{Carollo07,Carollo10} and \cite{Beers12}. Other works \citep[][]{deJong10,Kinman12,Kafle13,Hattori13,Chen14,Fernandez-Alvar15,Das16,Kafle17} that trace the more distant halo with giant stars or BHB stars also provide evidence for the duality of the halo.   
The two components exhibit distinct spatial density profiles, kinematics and metallicities.
For example, the inner halo demonstrates a flat density profile, and the nearly spherical profile in outer halo. 
\cite{Carollo07, Carollo10} provided the metallicity distribution functions (MDFs) of inner halo and outer halo which peaks at [Fe/H] $\sim$ $-1.6$ and [Fe/H] $\sim$ $-2.2$, respectively.   \cite{An15} used a sample of stars from SDSS to discover that the metallicity distribution of halo stars can be described by two Gaussian components
with peaks at [Fe/H] $\sim-1.4$ and $-1.9$ in the distance range 5 kpc $<d_{helio}<$ 10 kpc.   In addition,  the inner-halo stars, with distances up to 10-15 kpc from the Galactic center, exhibit a small net prograde rotation around the center of the Galaxy with $V_{\phi}$ $\sim$ 7 $\pm$ 4 km $s^{-1}$, and the outer halo stars which dominate in the region beyond $15-20$ kpc exhibit a retrograde net rotation with $V_{\phi}$ $\sim$ $-$80 $\pm$ 13 km $s^{-1}$. 
The difference in the characteristics of the components likely results from different formation mechanisms. 
Observations of spatial substructure in the Galaxy show that the outer halo was likely built purely via mergers \citep[][]{Newberg02, Belokurov06, Deason15}, while the formation process of the inner halo is still debated.  Some have suggested that the majority of the inner halo stars accreted in 10 Gyr ago, though some of them may formed in situ \citep{Norris94, Chiba00}. By studying the right structure of the nearby stellar halo with data from Gaia and RAVE, \cite{Helmi17} found that a high fraction of the local halo stars are on retrograde orbits, and provided the evidence that the inner halo was built entirely via accretion.

\par Studies of the detailed chemical abundances of halo stars have sought to place further constraints on the formation of the Galactic halo 
\citep{Nissen97, Stephens02, Nissen10, Feltzing13, Haywood13}.  
Because the $\rm{\alpha}$-element to iron abundance ratio $\rm{[\alpha/Fe]}$
reflects the nucleosynthesis and chemical enrichment history of a stellar population \citep{Tinsley79, Matteucci86}, the distribution of [$\alpha$/Fe] provides information about the birthplace of the stars. 
Most $\alpha-$elements are produced in the explosion of Type II supernovae, which increase element abundance in a timescale of $10^6-10^7$ years.  Type Ia supernovae produce primarily iron peak elements (especially Fe), and enrich the interstellar medium after a longer timescale ($\sim$ $10^9$ year) \citep{Maoz10}.  The [$\alpha$/Fe] abundance therefore indicates the timescale and environment in which a stellar population is born.

The [$\alpha$/Fe] abundance has been
shown to be very powerful disentangle the various Milky Way components, indicating that they have had different formation histories. 
Recent research found that the thick disk and the inner-halo stars have high-[$\alpha$/Fe] abundances \citep{Ishigaki12}. 
\cite{Nissen10} demonstrated that in a local sample of stars, those with high [$\alpha$/Fe] have a constant have a constant [$\alpha$/Fe] of $\sim0.3$ in the metallicity range $-2 <$ [Fe/H] $<-0.5$, while stars with lower [$\alpha$/Fe] of about 0.15 exhibit a decrease in [$\alpha$/Fe] toward higher metallicity;  the two halo populations can be separated based on measurements of [$\alpha$/Fe].  \cite{Carollo12,Carollo14} used Carbon-enhanced stars to show that the Galactic halo is composed of two stellar populations instead of one.  \cite{Hawkins15b}  discovered that the thick disk and halo are not chemically distinct in [$\alpha$/Fe], indicating a smooth
transition between thick disk and halo.  They analyzed the chemical distribution of stars with $-1.20<$ [Fe/H] $<-0.55$ in the SDSS \uppercase\expandafter{\romannumeral3}'s Apoche Point Observatory Galactic Evolution Experiment \citep[APOGEE:][]{Eisenstein11} to argue that the Galactic halo and thick-disk are formed from chemically similar gas, but one component is pressure-supported while the other is angular momentum-supported. 

\par A better understanding of the Galactic halo requires more information such as chemical abundance and kinematics of large number of individual stars. The ongoing Large Sky Area Multi-Object Fiber Spectroscopic Telescope (LAMOST) survey has released  more than 6 million stellar spectra with stellar parameters in the DR4 catalogue. This data set allows us to expand on previous studies of the halo and to constrain the structure and evolution of the Galaxy. 

\par In this study, we make use of  A/F/G/K-type giant stars selected from the LAMOST survey to explore the metallicity distribution and kinematics of the Galactic halo. The outline of this paper as follows. In Section 2, we present a brief overview of the LAMOST observational data, and list the selection criterion for our stellar sample. In Section 3 we present the observed metallicity distribution of halo stars.  Kinematics analysis of halo stars are presented in Section 4.  
A brief summary is given in Section 5.
 
\section{Data}

\par LAMOST is a large field multi-object telescope located at Xinglong Station, which is part of the National Astronomical Observatories, Chinese Academy of Sciences (NAOC). 
LAMOST is a reflecting Schmidt telescope with its optical axis fixed along the north-south meridian.  It can take 4000 spectra simultaneously in a single exposure \citep{Cui12,Deng12,Zhao12}.  Its effective aperture varies from 3.6 to 4.9 m in diameter, depending on the altitude and azimuth observed, and it has a large, 20 deg$^2$ field of view.
The LAMOST spectrographs have a resolution of R $\rm \sim$ 1,800 and the observed wavelength range spans 3,700 {\AA} $\sim$ 9,000 {\AA} \citep{Cui12}.  
The survey reaches a limiting magnitude of $r=17.8$ (where $r$ denotes magnitude in the SDSS $r$-band), but most targets are brighter than $r\sim17$.   
The LAMOST Spectroscopic Survey of the Galactic Anticentre (LSS-GAC) \citep{Xiang17b} adds distance, [${\rm{\alpha}}$/Fe] abundance ratio ([${\rm{\alpha}}$/M]), elemental abundances [C/H] and [N/H], and absolute magnitudes $\rm{M_V}$ and $\rm{M_{Ks}}$ to the LAMOST Data Release catalogues.   
 
\par The stellar parameters ([Fe/H], log(g)) are derived with the ULySS software by template matching with the MILES spectral library \citep{Sanchez06, Wu11}. The accuracy of the pipeline was tested by selecting 771 stars from the LAMOST commissioning data with spectra of relatively good quality, and comparing with the SDSS/SEGUE Stellar Parameter Pipeline (SSPP). The precisions of the log(g) and [Fe/H] were found to be 0.34 dex and 0.16 dex, respectively. The [$\alpha$/Fe] are derived by template matching with the Kurucz synthetic spectral library \citep{Li16, Xiang17b}. 
To provide a realistic error estimate for $\rm{[\alpha/Fe]}$, the random error induced by spectral noise is combined with the systematic error, which is assumed to have a constant value of 0.09 dex.  The systematic error was estimated by a comparison with high-resolution measurements \citep{Li16}. A cross-identification of the LSS-GAC catalogues with APOGEE giant stars with 3500 $< \rm{T_{eff}} <$ 5300 K and $\rm{log(g)} <$ 3.8  yields 3,533 common objects that have a SNR higher than 50 in the LAMOST spectrum.  The standard deviation of the residuals between these two catalogues are only 0.04 dex for $\rm{[\alpha/Fe]}$ (as shown in the upper right panel of Figure 8 of \cite{Xiang17a}). 

\begin{figure}
\includegraphics[width=1.0\hsize]{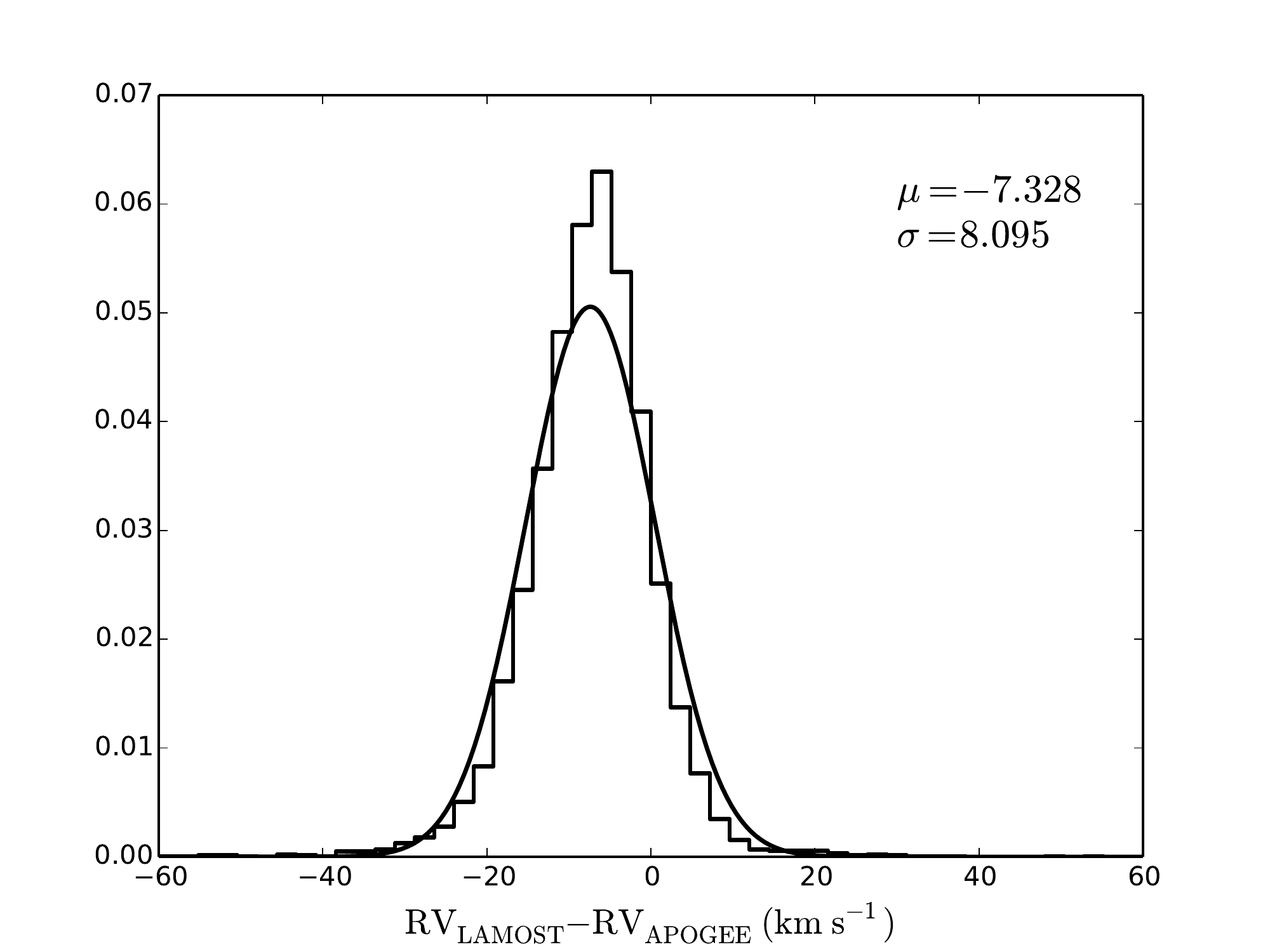}
\caption{Offsets of the radial velocities between LAMOST and APOGEE. The histogram shows the distribution of radial offset for 17,203 giant stars found in both surveys, and the solid curve is a Gaussian fit to this distribution.  Note that LAMOST spectra are systematically offset by -7.328 km s$^{-1}$ from the APOGEE measurements.}
\label{figure1}
\end{figure}

\begin{figure}[]
\includegraphics[width=1.0\hsize]{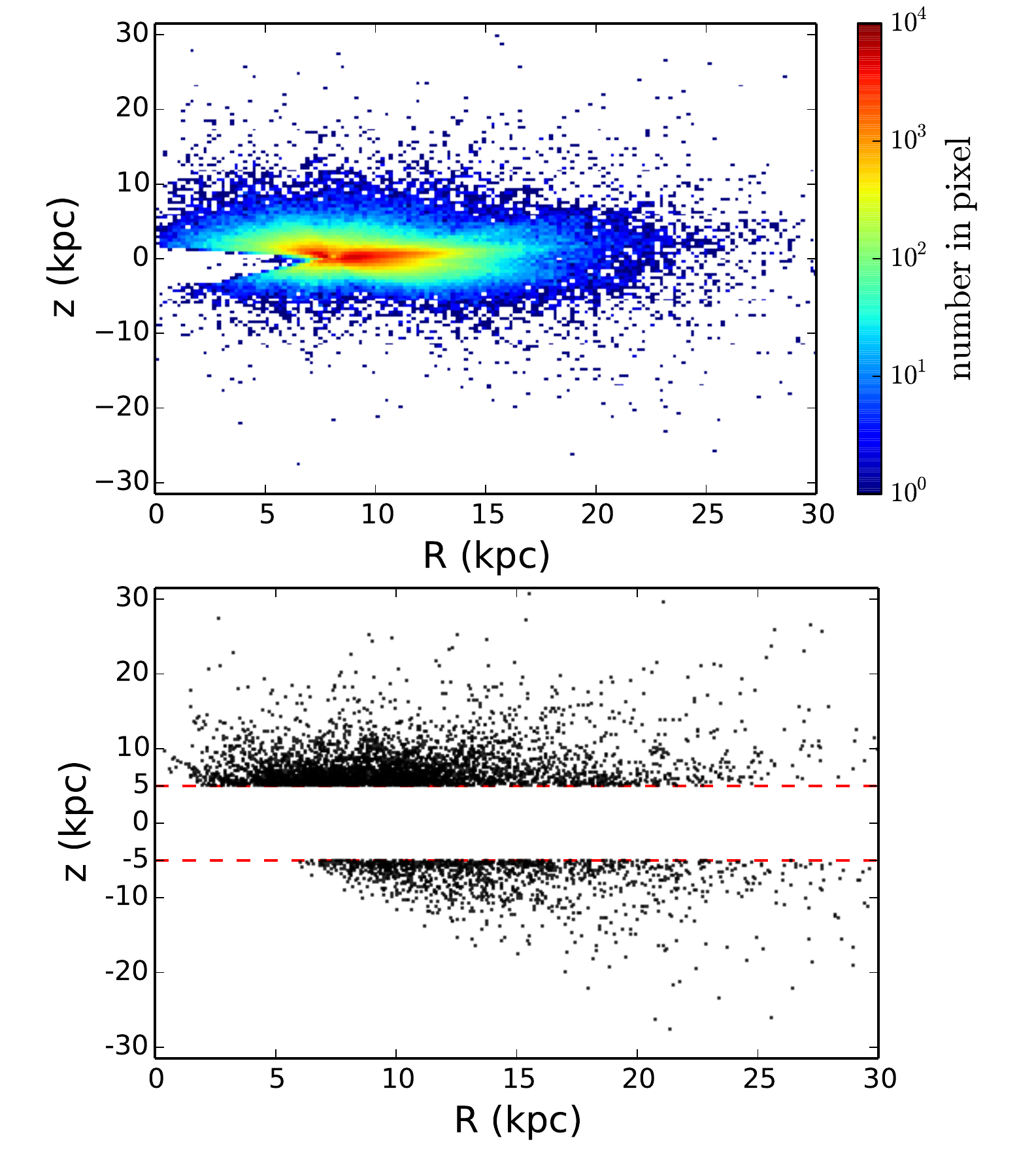}
\caption{The space distribution in cylindrical Galactic coordinates for the sample of 350,385 giant stars (upper panel), and 5,300 giant stars with $|z|>$5 kpc (lower panel). $Z$ is the distance from the plane of the Galaxy, and $R$ is the projected distance from the Galactic center in the Galactic plane. Note that most stars have $|z|<$20 kpc.}
\label{figure2}
\end{figure}

\par The data used in our work from two catalogues. The stellar parameters ([Fe/H], log(g)) and the line-of-sight velocity are from the LAMOST DR4 A/F/G/K-type stars catalog, and the proper motion and distance are from the LSS-GAC  catalog. There are 3,350,614 with measurements in both of these catalogs. 
In this work, we select giant stars by restricting
${\rm{0<log(g)<3.5}}$ and ${\rm{S/N>15}}$ in the $g$-band; we eliminate those stars without stellar atmospheric parameters (including spectral types, [Fe/H], [$\rm{\alpha/Fe}$], $\rm{log(g)}$, radial velocity, distance and proper motion). 

To check the systematics of the radial velocities, we match our data with the APOGEE data released in SDSS DR14 (average $\rm{S/N>15}$) and obtain 17,203 giant stars in common. Figure \ref{figure1} shows that the radial velocity offset between LAMOST DR4 stars and APOGEE stars in SDSS DR14 catalog is $-7.328$ km $\rm{s^{-1}}$ with a standard deviation is 8.095 km $\rm{s^{-1}}$.  \cite{Jing16} reports the offset in radial velocities between the LAMOST dwarfs and the SDSS-SSPP concentrate on $-6.76$ km~s$^{-1}$ with a dispersion of 7.9 km~s$^{-1}$.  \cite{Tian15} also finds that the radial velocity derived from the LAMOST pipeline is slower by $5.7$ km $\rm{s^{-1}}$ relative to APOGEE. In this paper, we reduce the LAMOST radial velocities to match the other surveys.

The proper motion from PPMXL catalog \citep{Roeser10}, combined with the tabulated distances and radial velocities, are used to derive the Galactocentric Cartesian velocity components of U, V, W, and Galactocentric cylindrical velocity components of $\rm{V_\phi}$, $\rm{V_R}$, $\rm{V_Z}$.  Here, we adopt a Local Standard of Rest velocity  $V_{LSR} = 220$ km $\mathrm{s^{-1}}$ \citep{Gunn79, Feast97} , and a Solar peculiar motion 
$\rm{(U_{\odot},V_{\odot},W_{\odot})}$ = $\rm{(-11.1,-12.24,7.25)\ km\ s^{-1}}$\citep{Schonrich10}, and $\rm{R_{\odot}}$ = 8 kpc.  

We add the criteria of the relative error of distance less than 0.4, the error of proper motion less than 10 mas yr$^{-1}$,and ($|\rm{V_\phi}|$, $|\rm{V_R}|$, $|\rm{V_Z}|$) $<$ 500 km $\mathrm{s^{-1}}$ (the escape speed of the Galaxy). The upper panel of Figure \ref{figure2} gives the spatial distribution in $R-Z$ plane for the sample of total 350,385 giant stars. As shown in Figure \ref{figure2}, most stars lie in the region $\rm{|z|\leqslant}$ 20 kpc.  
In order to derive possible halo sample stars, we restrict the $|z|\geqslant$5 kpc, as shown in the lower panel of Figure \ref{figure2}. There are 5,300 stars with $|z|\geqslant$ 5 kpc.

\begin{figure}
\includegraphics[width=1.0\hsize]{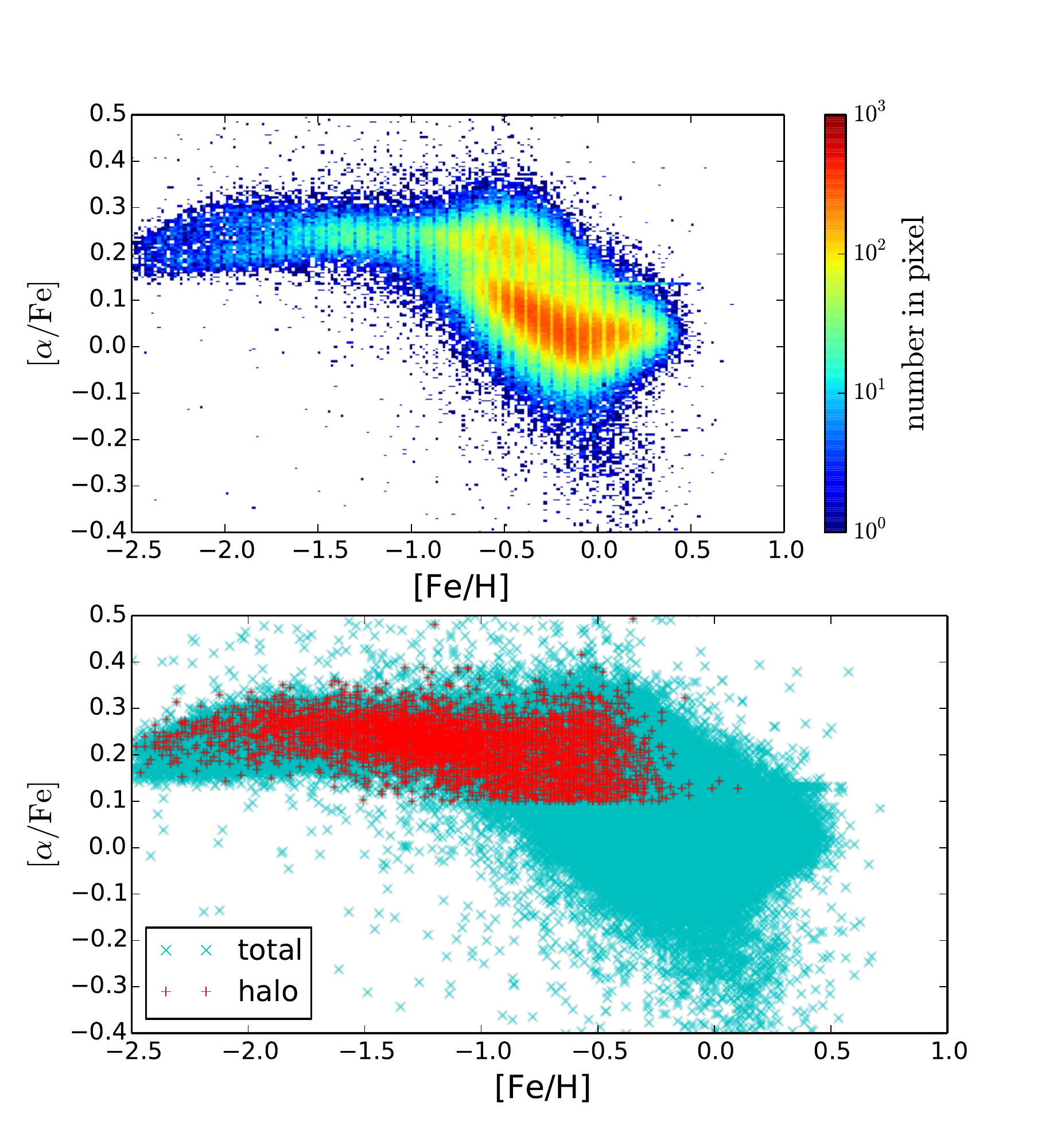}
\caption{Upper panel: Chemical abundance distribution, [$\alpha$/Fe] vs. [Fe/H], of sample stars; the color coding corresponds to the number density in each pixel.  
Lower panel: The distribution, [$\alpha$/Fe] vs. [Fe/H], for likely halo stars (red color) and all stars (blue color),  respectively.  }
\label{figure3}
\end{figure}

\section{The metallicity distribution of halo stars}

\par Galactic components have been distinguished in phase space, and by using [$\alpha$/Fe] to probe the Galactic components, using relatively large samples of high-resolution spectra \citep{Nissen10, Feltzing13, Haywood13}.  It has been shown that the $\rm{\alpha}$-element to iron abundance ratio $\rm{[\alpha/Fe]}$ is a good indicator of the Galactic chemical enrichment history \citep{Lee11}.  In our sample, the upper panel of Figure \ref{figure3} shows the chemical abundance distribution [$\alpha$/Fe] vs. [Fe/H].  
As shown in Figure \ref{figure3}, we found that the sample stars in the region [Fe/H]$>-1$ are divided to two sequence: 
a lower $\alpha$ sequence and an higher $\alpha$ sequence.  The two sequences correspond to 
the lower $\rm{[\alpha/Fe]}$ thin disk component and the higher $\rm{[\alpha/Fe]}$ thick-disk component \citep{Lee11}. 
The lower panel shows the distribution of $\rm{[\alpha/Fe]}$ vs. metallicity for the possible halo stars (red color) and total stars (blue color), respectively.

Comparing the upper panel and the lower panel of Figure \ref{figure3}, the halo stars have a similar distribution of $\rm{[\alpha/Fe]}$ ($\rm{0.1\leqslant [\alpha/Fe]\leqslant 0.3}$) to that of thick-disk stars within $\rm{-1.0<[Fe/H]<-0.5}$; most halo stars are concentrated in more metal-poor regions.  \cite{Hawkins15b} also discovered that there is no chemical distinction between canonical thick-disk and canonical halo stars in APOGEE data with $\rm{-1.20<[Fe/H]<-0.55}$. 
We notice that a small fraction of our sample stars have a lower value of [$\rm{\alpha}$/Fe], which indicates the presence of disk stars in our sample; we therefore remove the stars with both low $\alpha$-element abundance and higher metallicity ([$\rm{\alpha}$/Fe] $<$ 0.1 and [Fe/H]$>-1$), which reduces the sample to 4,850  halo stars.

\par
\begin{figure}[!hbp]
\includegraphics[width=1.0\hsize]{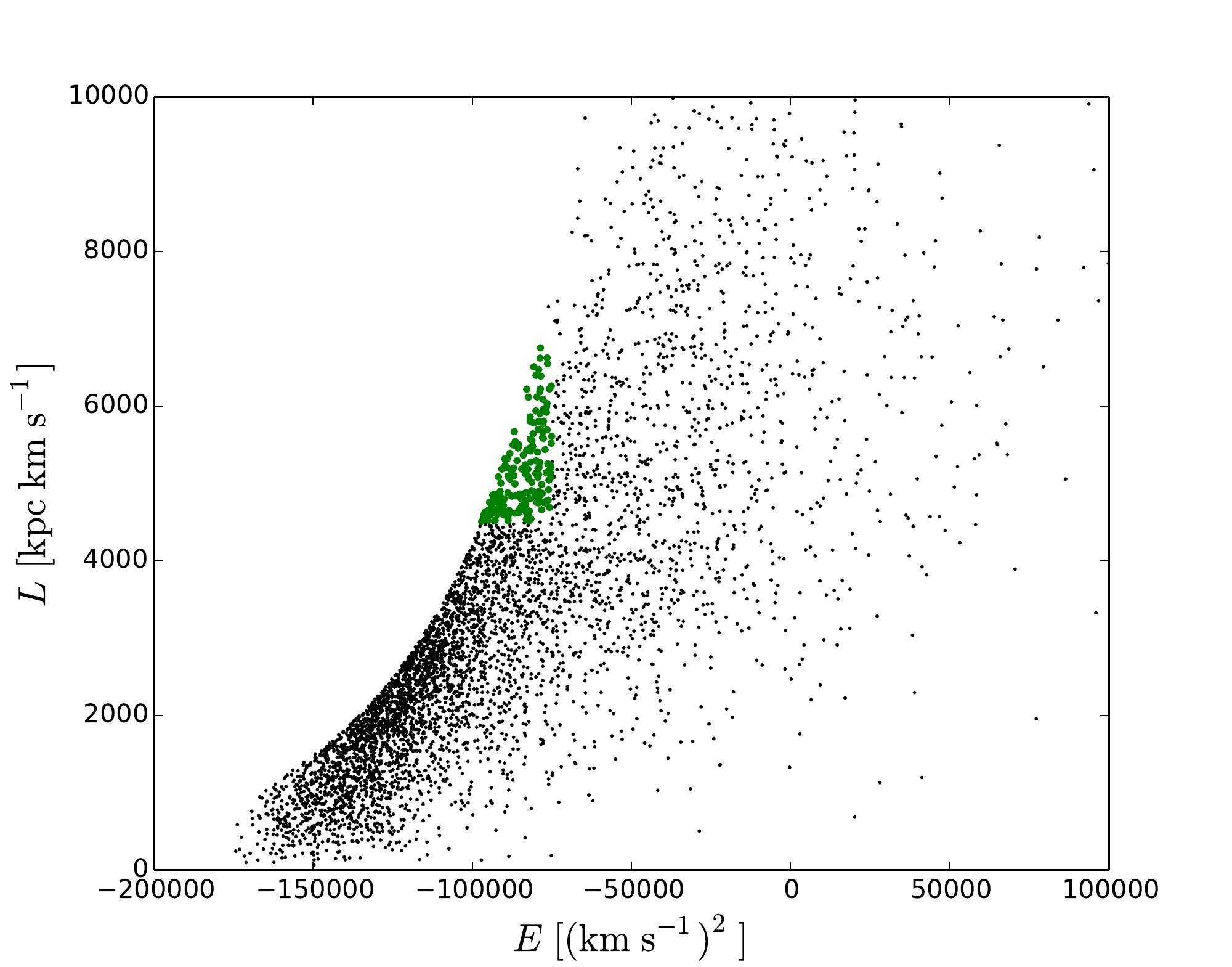}
\caption{Energy $E$ versus total angular momentum $L$ for the full sample of 4850 halo stars (black points). Stars identified with the Sagittarius stream have been marked in green.}
\label{figure4}
\end{figure}

 \textbf{ In order to remove the influence of Sagittarius stream on our results, we have tagged Sagittarius in the distribution of energy $E$ versus total angular momentum $L$. We adopt the gravitational potential model provided by \cite{Xue08}, which includes three components: a spherical \citet{Hernquist90} bulge, an exponential disk, and an NFW dark matter halo \citep{Navarro96}.The energy $E$ is derived by gravitational potential and total velocity:
\begin{align}
E=\Phi+ V_{tot}^2/2  \nonumber
\end{align}
where $V_{tot}^2=U^2+V^2+W^2=V_{r}^2+V_{\theta}^2+V_{\phi}^2$. The total angular momentum $L$:
\begin{align}
L=r_{gc}V_{tan}  \nonumber
\end{align}
where $r_{gc}=\sqrt{X^2+Y^2+Z^2}$ and $V_{tan}=V_{\theta}^2+V_{\phi}^2$.
Sagittarius stream member stars are primarily found at $E=-90000$ $(\rm{km\ s^{-1}})^2$ and $L$ = 6000 kpc $\rm{km\ s^{-1}}$ \citep{Bird18}. As shown in Figure \ref{figure4},  we mark these stars in green points and there are 170 giant stars. We remove them from the sample and remain 4,680 halo stars.}

\par
\begin{figure}
\includegraphics[width=1.0\hsize]{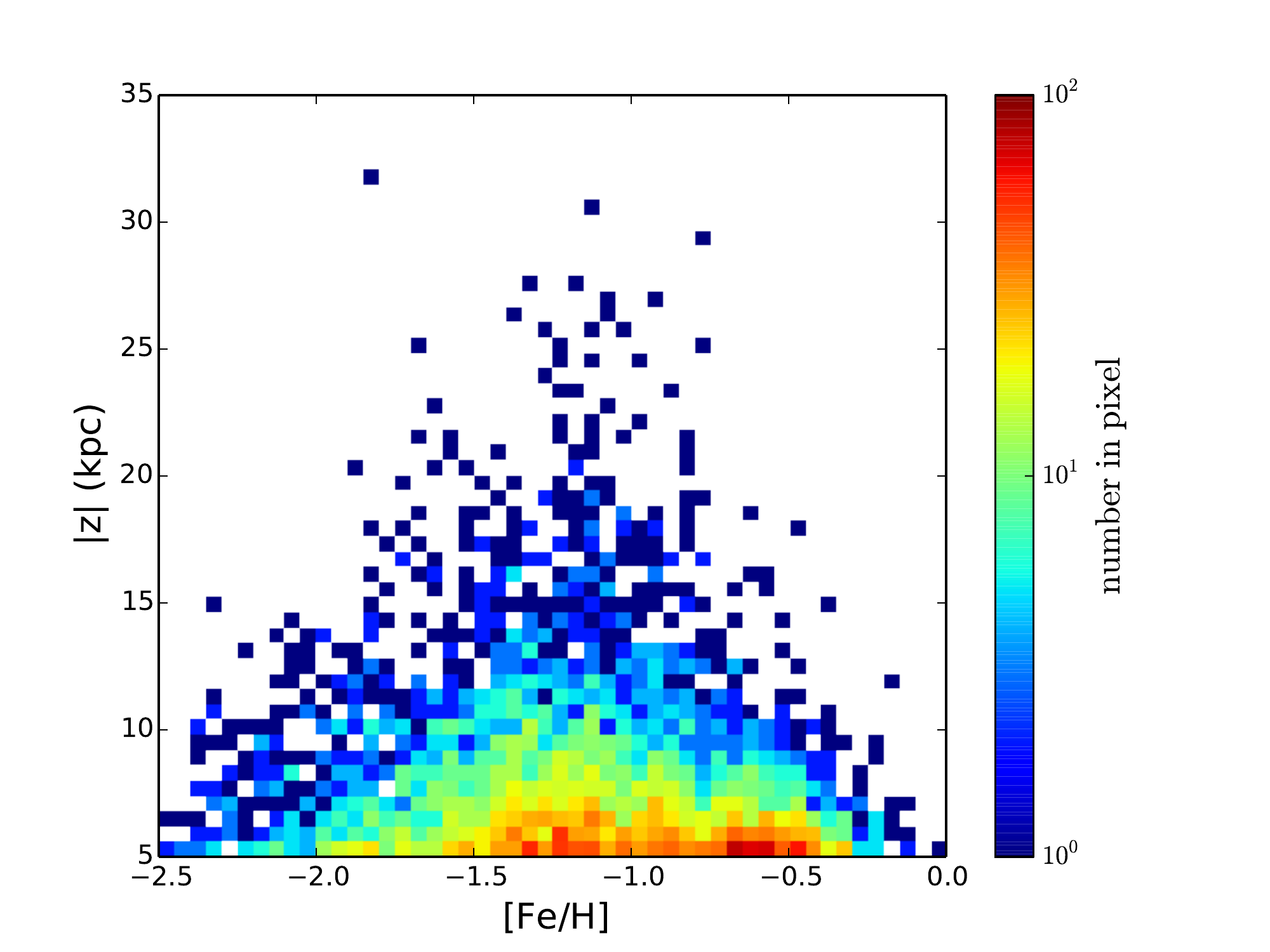}

\caption{The distribution of vertical distance $|z|$ versus metallicity 
[Fe/H]; the color coding corresponds to the number density in each pixel. }
\label{figure5}
\end{figure}

 Figure \ref{figure5} shows vertical distance $|z|$ versus metallicity 
[Fe/H] in our final sample of primarily halo stars. 
The figure shows that metallicity is concentrated in two peaks: one located between
[Fe/H] $\sim-$0.5 and $\sim-$0.8, and the second between [Fe/H] $\sim-$1 and $\sim-$1.3.  
The higher metallicity peak probably represents a small percentage of thick disk stars that remain in the sample despite the distance cut ($|z| > 5$ kpc) and the selection based on $\alpha$-elements ($\rm{[\alpha/Fe]}\leqslant -1$).  
It is difficult to derive a pure halo sample stars.
In addition, our sample is small enough that it does not include representatives of very metal-poor stars ([Fe/H] $<-2.0$), especially for $|z|>$15 kpc. Therefore, we restrict our study of the metallicity distribution of halo stars to the region 5 kpc $<|z|\leqslant$ 15 kpc.   
In order to explore the relationship between MDFs and distance from the Galactic plane, we divide the sample stars into different distance intervals.  Assuming the MDFs of stars from a single stellar population are expressed by a single Gaussian function,  we fit the observed MDFs with Gaussian mixture model and derive the optimal parameters of each component \citep{Zuo17}.  

\begin{figure}[!hbp]
\includegraphics[width=1.0\hsize]{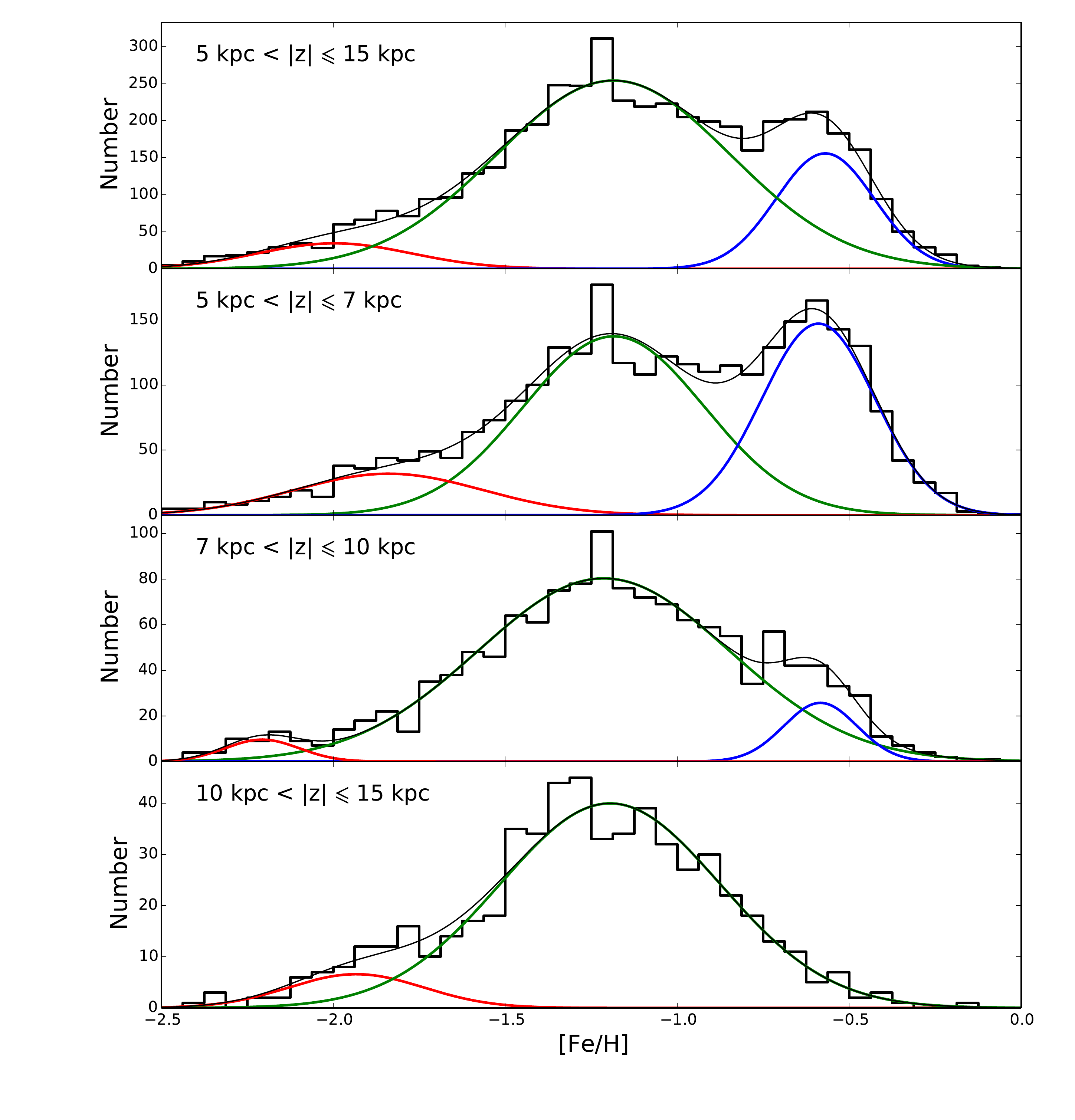}
\caption{The metallicity distribution of sample stars in different intervals of distance from the Galactic plane. The red, green and blue curves represent components associated with the outer-halo, the inner-halo, and the thick-disk components, and their sum is illustrated by the black curve.}
\label{figure6}
\end{figure}

We determine the optimal number of Gaussian functions using the Bayesian information criterion ($BIC$):
\begin{align}
BIC=-2ln[L^{0}(M)] + klnN  \nonumber
\end{align}
where $L^{0}(M)$ represents the maximum value of the likelihood function of the model, $N$ is the number of data points, and $k$ represents the number of free parameters. More information about the $BIC$ can be found in \cite{Ivezic14}. Figure \ref{figure6} shows the lowest $BIC$ fits to the data: a three-peak, three-peak, three-peak and two-peak Gaussian mixture models fit to MDFs with 5 kpc $<|z| \leqslant$ 15 kpc, 5 kpc $<|z|\leqslant$ 7 kpc, 7 kpc $<|z|\leqslant$ 10 kpc and 10 kpc $<|z|\leqslant$ 15 kpc, respectively.  We infer that blue, green and red curves represent the thick-disk, inner-halo and outer-halo components, with peaks at [Fe/H] $\sim-0.6\pm0.1$, $-1.2\pm0.3$ and $-2.0\pm0.2$, respectively, in the distance of 5 kpc $<|z| \leqslant$ 15 kpc.

The upper panel in Figure \ref{figure6} shows that there are three components in the distance interval, and the inner-halo component occupies the vast majority. The lower three panels in Figure \ref{figure6} illustrate the change in the fraction of each component with increasing distance $|z|$.
Table \ref{Table 1} gives the detailed results of the mean metallicites, weights and full width at half maximum (FWHM) of the three components in different vertical distance intervals. These results show that the proportions of thick disk decline from 35$\%$ to nearly 0, so that the bin farthest from the Galactic plane can be fit with only two components. The fraction of stars in the inner halo component rises from 52$\%$ to 91$\%$ with increasing distance from the plane. However, for the outer halo component, it shows that there is no clear correlation between the proportion in this component and increasing distance, due to the incompleteness of very metal-poor stars at large distances from the plane in our sample.

Note that the result in Figure 20 of \cite{Carollo10} shows the shift of metallicity peak from $-0.6$ at 0 kpc$<|z|<$ 1 kpc to $-1.3$ at 3 kpc$<|z|<$ 4 kpc, reaching to $-1.6$ at 6 kpc$<|z|<$ 7 kpc and $-2.2$ at $|z|>9$ kpc.    In our work,  the inner halo and outer peak at [Fe/H] = $-1.2\pm0.3$ and $-2.0\pm0.2$.  As compared with the results which peak at [Fe/H] = $-1.6$ and $-2.2$ from MS stars with 4 kpc of the sun in the SDSS survey by \cite{Carollo07, Carollo10}, our result is higher than their values.
However,  some other studies also give different results.  For example,  \cite{An15} used a sample of main-sequence stars from SDSS and discover that the metallicity distribution of halo stars peak at [Fe/H] $\sim-1.4$ and $-1.9$ in the distance range 5 kpc $<d<$ 10 kpc.  
\cite{Zuo17} used F/G main-sequence turnoff stars in the south Galactic cap to found that the halo stars peak at [Fe/H]$\sim-0.63$, $-1.45$, and $-2.0$.  \cite{Chen10} used a large sample of red horizontal-branch to determine a two peak metallicity distributions, one peaking at [Fe/H] $\sim-0.6$, mainly originates from the thick disk, and another peaking at [Fe/H] $\sim-1.3$, generally belongs to the halo.  Therefore, the exact metallicity peaks at a given $|z|$ between the present work and those references are somewhat different due to sample selection of different spectral types of stars.

\begin{table*}[ht]
\centering
\caption{ {\upshape The mean metallicites, weights and full width at half maximum(FWHM) of thick-disk, inner-halo and outer-halo in different distance intervals}}
\label{Table 1}
\begin{tabular}{cccccccccc}
\hline
\hline 
Distances& \multicolumn{3}{c}{Thick-disk} & \multicolumn{3}{c}{Inner-halo} & \multicolumn{3}{c}{Outer-halo} \\

(kpc) & Mean  & Weight  & FWHM & Mean & Weight & FWHM & Mean & Weight & FWHM \\\hline
$\rm{5<|z|\leqslant15}$ & $-0.6$ & 19$\%$ & 0.3 & $-1.2$ & 74$\%$ & 0.8 & $-2.0$ & 7$\%$ & 0.5\\

$\rm{5<|z|\leqslant7}$ & $-0.6$ & 35$\%$ & 0.4 & $-1.2$ & 52$\%$ & 0.6 & $-1.8$ & 13$\%$ & 0.6\\

$\rm{7<|z|\leqslant10}$ & $-0.6$ & 8$\%$ & 0.3 & $-1.2$ & 89$\%$ &  0.9 & $-2.2$ & 3$\%$ & 0.3 \\

$\rm{10<|z|\leqslant15}$ &  &  & & $-1.2$ & 91$\%$ & 0.8 & $-1.9$ & 9$\%$ & 0.5\\
\hline
\end{tabular}
\end{table*}

\section{Kinematics of halo stars}
\subsection{Correlation of orbital eccentricities with metallicity and distance}

\par
\begin{figure*}
\centering
\includegraphics[width=1.0\textwidth]{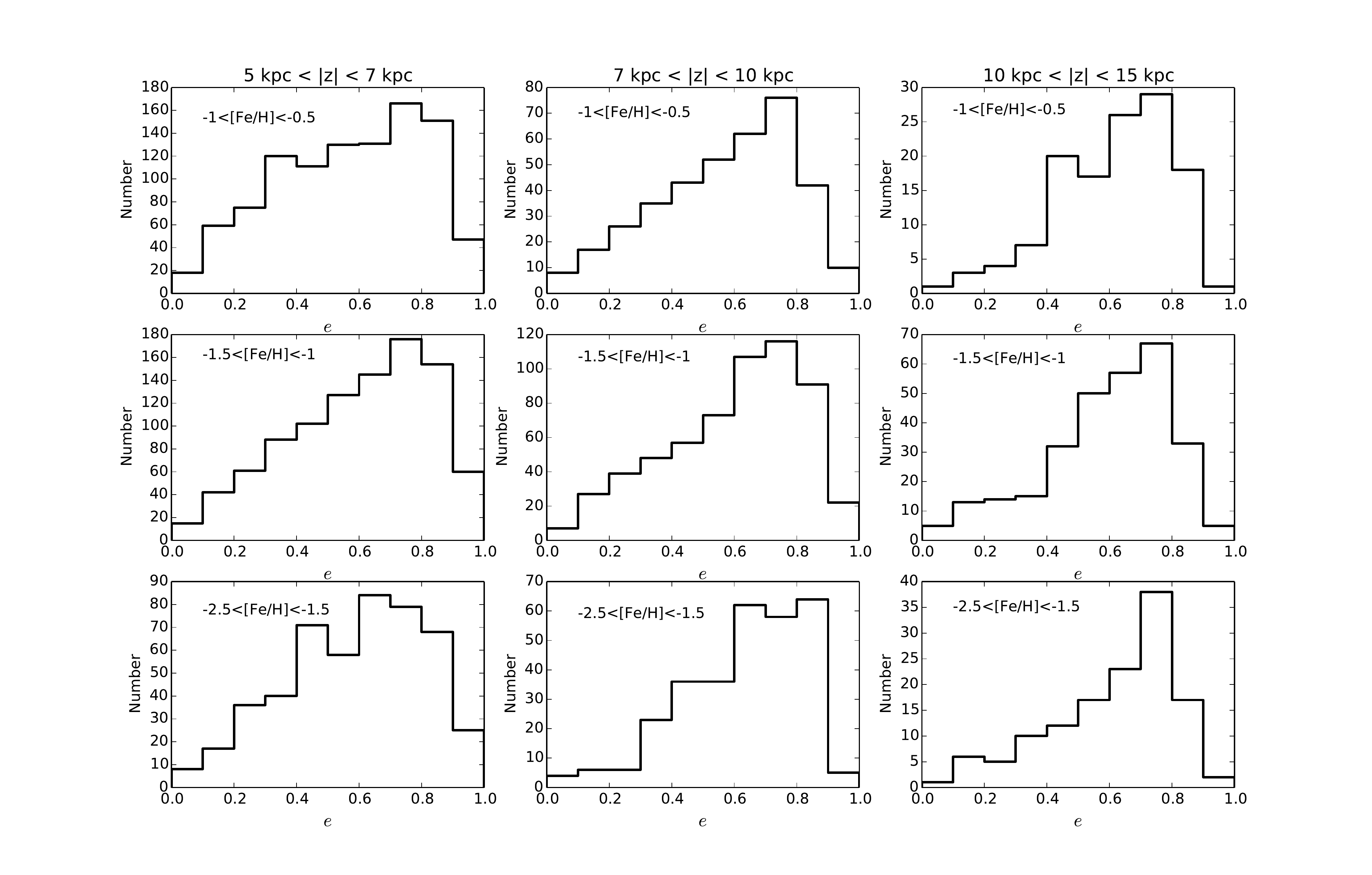}
\caption{Distribution of the orbital eccentricity in different metallicity ranges and different intervals on distance 
 $|z|$ from the Galactic plane of 5 kpc $<\ |z|\ <$ 7 kpc, 7 kpc $<\ |z|\ <$ 10 kpc, and 10 kpc $<\ |z|\ <$ 15 kpc, respectively.}
\label{figure7}
\end{figure*}

 Based on the gravitational potential model proposed by \cite{Xue08}, we compute the orbital eccentricity, $e$, defined as
$e=(\rm{r_{apo}}$ - $\rm{r_{peri}}$ )/($\rm{r_{apo}}$ + $\rm{r_{peri}}$), where $\rm{r_{apo}}$ and $\rm{r_{peri}}$ represent the farthest and the closest extent of an orbit from the Galactic centre, respectively. Errors in the derived orbital parameters for each star are primarily due to errors in the distance and proper motion.
Figure \ref{figure7} shows the distribution of derived orbital eccentricity for stars in different metallicity ranges and different intervals on distance 
 $|z|$ from the Galactic plane of 5 kpc $<\ |z|\ <$ 7 kpc, 7 kpc $<\ |z|\ <$ 10 kpc and 10 kpc $<\ |z|\ <$ 15 kpc, respectively.  For stars with $-1.0 <$ [Fe/H] $<-$0.5,  the canonical thick disk stars exist in the range of 5 kpc $<\ |z|\ <$ 7 kpc, and the ratio of $e<$ 0.4 declines with increasing vertical distance and decreasing metallicty.  Farther from  the Galactic plane the observed distribution of eccentricity is more consistent with the inner-halo stars, and the peak values maintain $e\sim$ 0.75 in most panels.  But the peak values of lower left two panels,  for metal-poor halo stars with $-2.5 <$ [Fe/H] $< -$1.5, change to $e\sim$ 0.65, $e\sim$ 0.85, in vertical distance of 5 kpc $<\ |z|\ <$ 7 kpc and 7 kpc $<\ |z|\ <$ 10 kpc.   We consider this distribution in the metal-poor regions reflects the contribution from an outer-halo population with a fraction of stars on high eccentric orbits, superposed on the inner-halo population. 
The Figure 5 of \cite{Carollo10} shows the similar distribution of orbital eccentricity in two different vertical distance interval of 1 kpc $<\ |z|\ <$ 2 kpc and 2 kpc $<\ |z|\ <$ 4 kpc, and it also shows a linear eccentricity distribution for the metal-poor halo stars with $-2.0 <$ [Fe/H] $<-1.5$.   But the approximate linear distribution of eccentricity appears not prominent in our distribution.  Comparing to their study,  our sample stars have higher vertical distance, so the thick disk stars have less influence and our sample stars are dominated by halo stars.

\subsection{Correlation of rotational velocity with metallicity}
\par
\begin{figure*}
\centering
\includegraphics[width=1.0\textwidth]{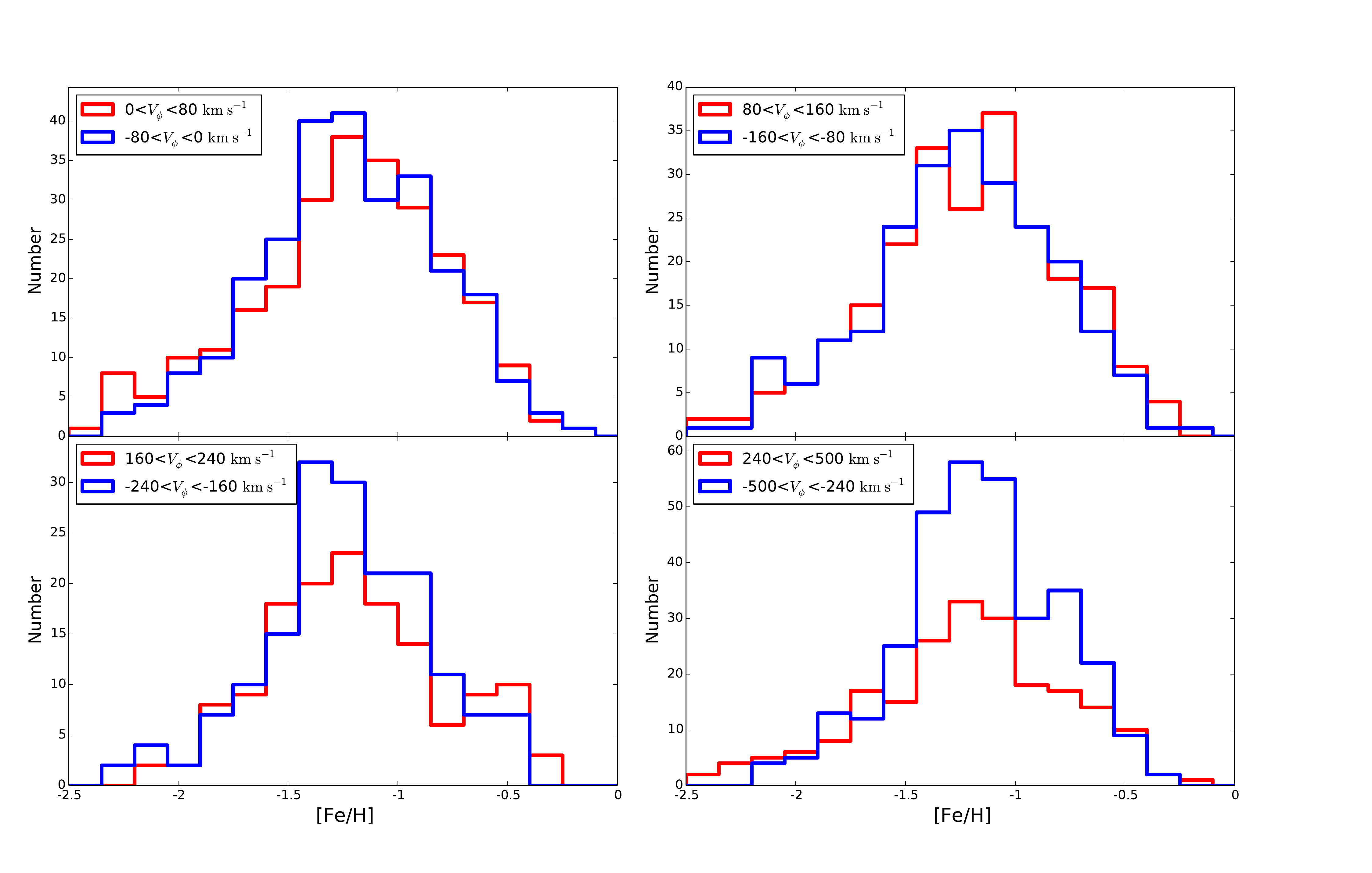}
\caption{The probability distributions of the metallicity for sample stars within  $7 < |z| <15$ kpc, and rotation velocity ($|v_{\phi}|$) in the range 0-80 $\rm{km\ s^{-1}}$ ($top$ $left$; 519 stars), 80-160 $\rm{km\ s^{-1}}$ ($top$ $right$; 454 stars), 160-240 $\rm{km\ s^{-1}}$ ($bottom$ $left$; 311 stars), 240-500 $\rm{km\ s^{-1}}$ ($bottom$ $right$; 527 stars). Stars with prograde velocity are represented by the red histogram and retrograde velocity are  the blue histogram.}
\label{figure8}
\end{figure*}

 In this section, we study the correlation of rotational velocity with metallicity, [Fe/H]. Stars with v$_{\phi} > 0$ km s$^{-1}$ are prograde, while stars with v$_{\phi} < 0$ $\rm{km\ s^{-1}}$ are retrograde. The Galactic disk has a rotation velocity of v$_{\phi}$ = $220$ km s$^{-1}$, and is thus prograde.  To eliminate the influence of the thick-disk stars and obtain a purely halo sample, we selected only the 1,889 stars in the sample with 7 $<|z| \leqslant$ 15 kpc. We divide these stars into four velocity intervals.
Figure \ref{figure8} shows the MDFs in different $\rm{v_{\phi}}$ intervals of prograde and retrograde velocities; there are 519, 454, 311 and 527 stars in $|$v$_\phi|$ intervals, [0, 80], [80, 160], [160, 240] and [240, 500], respectively. 
The number of retrograde stars, which are represented by the blue histogram, is 264, 224, 169 and 319 in each panel, from low to high angular speed. 

As seen in this figure, the majority of the most metal-poor stars are retrograde.  \cite{Zuo17} also derived a similar result based on SCUSS and SDSS data.
However, since they lacked radial velocity measurements, they only selected higher Galactic latitude ($b < -45^{\circ}$) stars, since the rotational speeds of these stars were only related to the proper motions, and relied much less on line-of-sight velocity measurements.  In comparison to their work,  our results have more distinct distributions of prograde and retrograde components in the halo sample. 

\section{Summary}

\par Based on 4,680 giant stars from the LAMOST DR4 stars, combined with measurements of $\rm{[\alpha/Fe]}$, proper motion and distance from the LSS-GAC catalog, we investigate the metallicity distributions and kinematics of stars with $|z|>$ 5 kpc.  $\rm{[\alpha/Fe]}$ and $z_{max}$ are used to distinguish various components of the Galaxy.  
The distribution of $\rm{[\alpha/Fe]}$ ($\rm{0.1\leqslant [\alpha/Fe]\leqslant 0.3}$) is similar for the thick-disk and halo stars 
within the metallicity range $\rm{-1.0<[Fe/H]<-0.5}$; most halo stars are concentrated in metal-poor regions.  
To eliminate thin-disk and thick-disk stars from our sample,  we remove the stars with [$\rm{\alpha}$/Fe] $<$ 0.1. In addition, we select only stars with $z_{max }>5$ kpc to increase the purity of the halo sample.  Then, we study the metallicity distribution of the sample stars within 5 $<|z|\leqslant$ 15 kpc.  

The metallicity distribution can be described by three-peak Gaussian model with peaks at [Fe/H] $\sim-0.6\pm0.1$, $-1.2\pm0.3$ and $-2.0\pm0.2$, and weights of 19$\%$, 74$\%$, and 7$\%$, corresponding to the thick disk, inner-halo, and outer-halo, respectively.  We also derive the mean metallicites and weights of three components in different distance intervals, and show that the proportions of thick disk stars in the sample decline from 35$\%$ to nearly 0$\%$; the metallicity distribution is well fit with only two components when $|z|>$ 10 kpc. For the inner halo component, the proportions rise from 52$\%$ to 91$\%$ with the increasing distance. However, for the outer halo component, our results show that there is no clear correlation between the proportion and the vertical distance, due to the incompleteness of our halo stars in very metal-poor and distant regions.
\par  To better understand the kinematics of the sample stars, 
we derive their orbital parameters by adopting a gravitational potential model.  According to the orbital eccentricity distribution in different metallicity ranges and different distance intervals , we notice that  the canonical thick disk stars exist in the range of 5 kpc $<\ |z|\ <$ 7 kpc and  $-1.0 <$ [Fe/H] $<-$0.5, and the ratio of $e<$ 0.4 declines with increasing vertical distance and decreasing metallicty.  Farther from  the Galactic plane the observed distribution of eccentricity is more consistent with the inner-halo stars, and the peak values maintain $e\sim$ 0.75,  independent of height above the plane, within the range 5$<|z|<$ 15 kpc.
Also, we study the correlation of rotational velocity with metallicity for sample stars within  $7 < |z| <15$ kpc. By comparing the MDFs in different rotation velocity intervals, we find that the majority of the retrograde stars are more metal-poor than the prograde stars.

\section*{Acknowledgements}

\par We thank especially the referee for insightful comments and suggestions, which have improved the paper significantly. 
This work was supported by joint funding for Astronomy by the National Natural Science Foundation of China and the Chinese Academy of Science, under Grants U1231113.  This work was also by supported by the Special funds of cooperation between the Institute and the University of the Chinese Academy of Sciences, and China Scholarship Council (CSC). 
HJN acknowledges funding from NSF grant AST-1615688. Funding for SDSS-III has been provided by the Alfred P. Sloan Foundation, the Participating Institutions, the National Science Foundation, and the U.S. Department of Energy Office of Science. This project was developed in part at the 2016 NYC Gaia Sprint, hosted by the Center for Computational Astrophysics at the Simons Foundation in New York City. 
In addition, this work was supported by the National Natural Foundation of China (NSFC, No.11373033, No.11373035, No.11625313, No.11573035), and by the National Basic Research Program of China (973 Program) (No. 2014CB845702, No.2014CB845704, No.2013CB834902). The Guoshoujing Telescope (LAMOST) is a National Major Scientific Project has been provided by the National Development and Reform Commission. LAMOST is operated and managed by the National Astronomical Observatories, Chinese Academy of Sciences.

\end{document}